\documentclass[sigconf]{acmart}
\AtBeginDocument{%
  }

\setcopyright{acmlicensed}
\copyrightyear{2025}
\acmYear{2025}
\setcopyright{rightsretained}
\acmConference[UIST Adjunct '25]{The 38th Annual ACM Symposium on User Interface Software and Technology}{September 28-October 1, 2025}{Busan, Republic of Korea}
\acmBooktitle{The 38th Annual ACM Symposium on User Interface Software and Technology (UIST Adjunct '25), September 28-October 1, 2025, Busan, Republic of Korea}\acmDOI{10.1145/3746058.3758982}
\acmISBN{979-8-4007-2036-9/2025/09}

\begin{document}
\title{QueryGenie: Making LLM-Based Database Querying Transparent and Controllable}

\author{Longfei Chen}
\email{chenlf@shanghaitech.edu.cn}
\orcid{0009-0002-4596-8093}
\affiliation{%
  \institution{ShanghaiTech University}
  \city{Shanghai}
  \country{China}
}

\author{Shenghan Gao}
\email{gaoshh1@shanghaitech.edu.cn}
\orcid{0009-0008-3397-0341}
\affiliation{%
  \institution{ShanghaiTech University}
  \city{Shanghai}
  \country{China}
}

\author{Shiwei Wang}
\email{three.wang@huolala.cn}
\orcid{0009-0004-0833-0896}
\affiliation{%
  \institution{Shenzhen Yishi Huolala Technology Co., Ltd}
  \city{Shenzhen}
  \country{China}
}

\author{Ken Lin}
\email{adam.lin@huolala.cn}
\orcid{0009-0002-5556-2572}
\affiliation{%
  \institution{Shenzhen Yishi Huolala Technology Co., Ltd}
  \city{Shenzhen}
  \country{China}
}

\author{Yun Wang}
\email{wangyun@microsoft.com}
\orcid{0000-0003-0468-4043}
\affiliation{%
  \institution{MircosoftResearch}
  \city{Hong Kong}
  \country{China}
}

\author{Quan Li}
\email{liquan@shanghaitech.edu.cn}
\orcid{0000-0003-2249-0728}
\affiliation{%
  \institution{ShanghaiTech University}
  \city{Shanghai}
  \country{China}
}
\authornote{Quan Li is the corresponding author.}

\begin{abstract}
  Conversational user interfaces powered by large language models (LLMs) have significantly lowered the technical barriers to database querying. However, existing tools still encounter several challenges, such as misinterpretation of user intent, generation of hallucinated content, and the absence of effective mechanisms for human feedback—all of which undermine their reliability and practical utility. To address these issues and promote a more transparent and controllable querying experience, we proposed \textit{QueryGenie}, an interactive system that enables users to monitor, understand, and guide the LLM-driven query generation process. Through incremental reasoning, real-time validation, and responsive interaction mechanisms, users can iteratively refine query logic and ensure alignment with their intent.
\end{abstract}



\maketitle

\section{Introduction}
\par Conversational user interfaces powered by large language models (LLMs), such as \textit{Vanna}~\cite{Vanna} and \textit{DataLine}~\cite{DataLine}, have significantly lowered the technical barriers to data querying, allowing non-expert users to access and explore databases with ease~\cite{li2024can,hong2024next}. By removing the need to write complex query languages or operate specialized tools, these systems enable users to interact with data more flexibly through natural language.

\par Despite their great potential, LLM-powered data querying tools still face significant reliability challenges, such as generating hallucinated content~\cite{xu2024hallucination}, misinterpreting user intent~\cite{gan2021towards,liu2023we}, or misunderstanding database schemas~\cite{li2024can}. These issues often lead to inaccurate query results and, in turn, undermine user trust. Most existing research~\cite{chang2023prompt,pourreza2024din,zhuang2024structlm} focuses on addressing these problems by improving model performance—for example, through prompt engineering or fine-tuning to optimize output quality.

\par  However, solely relying on model improvements is not sufficient to address users' confusion and uncertainty during real-world interactions. We argue that enabling users to see how the LLM reasons—understanding why and how a query is generated—is key to enhancing system transparency and building trust.

\par To this end, we developed \textit{QueryGenie}, an interactive system designed to help users understand and intervene in the query generation process of LLMs. The system follows a three-stage framework: Intention Confirmation – Query Generation – Query Validation. First, the system visualizes the mapping between the user's natural language input and the database schema, revealing how the LLM interprets user intent. Users can inspect and adjust this mapping to ensure that their intent is accurately captured before the system proceeds to generate structured query language (SQL). The query generation process adopts a chain-of-thought (CoT) approach, breaking down the query into a sequence of simpler, interpretable steps. Each step is accompanied by an explanation and an executable SQL statement, allowing users to examine and validate the logic in real time. If issues are detected, users can directly edit the SQL or provide feedback to the LLM for revision.

\begin{figure*}[h]
    \centering
    \includegraphics[width=\linewidth]{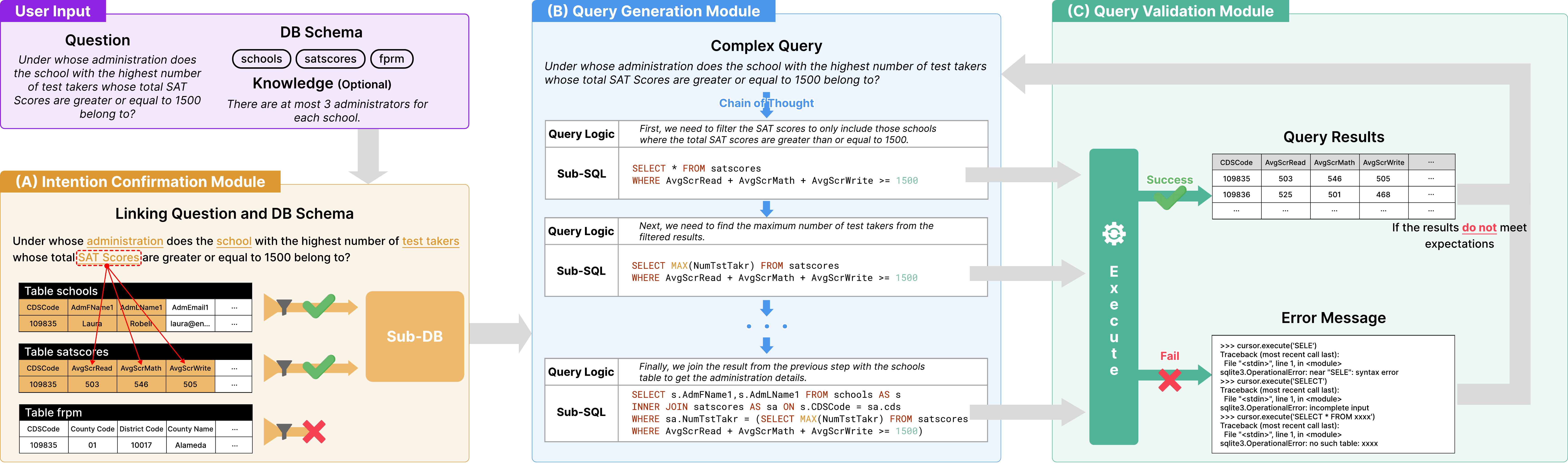}
    \vspace{-6mm}
    \caption{The framework of \textit{QueryGenie} consists of three key modules: (A) Intention Confirmation Module, which interprets user intentions from natural language input and maps them to the database schema; (B) Query Generation Module, which constructs query logic and converts it into precise SQL statements; and (C) Query Validation Module, which executes the SQL statements and returns the results.}
    \label{fig:framework}
    \vspace{-3mm}
\end{figure*}

\section{System Framework}
\par We propose a novel human–AI collaboration framework (\autoref{fig:framework}) aimed at enabling more controllable and transparent data querying.

\textbf{User Input.} To facilitate the subsequent discussion, we define the user input as a triplet $\mathcal{X=(Q,S,K)}$, where $\mathcal{Q}$ denotes the query question, $\mathcal{S}$ denotes the database schema, and $\mathcal{K}$ denotes external knowledge.

\textbf{Intention Confirmation Module.} This module verifies whether the LLM can accurately understand the user's intended query expressed in natural language. As shown in \autoref{fig:framework} (A), the system first guides the LLM to map entities in the query $\mathcal{Q}$ (e.g., ``SAT Scores'') with corresponding fields in the database schema $\mathcal{S}$ (e.g., \texttt{AvgScrRead}, \texttt{AvgScrMath} and \texttt{AvgScrWrite})—a process known as schema linking~\cite{lei2020re}. Users then review and adjust these mappings to ensure semantic alignment. If any mappings are incorrect, users can manually correct them or modify the input to trigger a new round of linking. Once accurate mappings between $\mathcal{Q}$ and $\mathcal{S}$ are established, the system filters out irrelevant portions of the schema, retaining only a focused subset $\mathcal{S'}$ for subsequent steps. This approach not only reduces the likelihood of irrelevant information affecting the query generation but also shortens the prompt length, thereby improving reasoning efficiency.

\textbf{Query Generation Module.} This module enhances the LLM's reasoning ability by using the CoT method to break down complex queries into a series of intermediate steps. As shown in \autoref{fig:framework} (B), the module guides the LLM to decompose complex queries into simpler sub-queries, with each sub-query corresponding to a specific sub-SQL statement. This approach not only helps users better understand the query logic but also improves the transparency of the generation.

\textbf{Query Validation Module.} This module works in conjunction with the Query Generation Module to individually check and verify each sub-query in the reasoning chain. As shown in \autoref{fig:framework} (C), the module executes SQL statements within the database and provides real-time results. This allows users to immediately observe the effects of each sub-query and quickly identify and correct any potential errors. If the result of a SQL query at a particular step does not meet expectations, users can provide immediate feedback by modifying the query logic or directly editing the SQL statements. The system will then update the subsequent steps in the reasoning chain accordingly. This incremental validation approach effectively mitigates the issue of error propagation in complex queries. By reviewing and adjusting each sub-query generation step, users can swiftly identify and rectify logical errors, preventing issues from becoming difficult to trace in the final SQL output.

\section{QueryGenie}
\par Based on the proposed framework, we developed a demo system called \textit{QueryGenie}, which integrates a database environment with an enhanced conversational interface to help users collaborate more effectively with LLMs in completing data querying tasks.

\subsection{Database Panel}
\par
\begin{figure*}[h]
    \centering
    \includegraphics[width=0.88\textwidth]{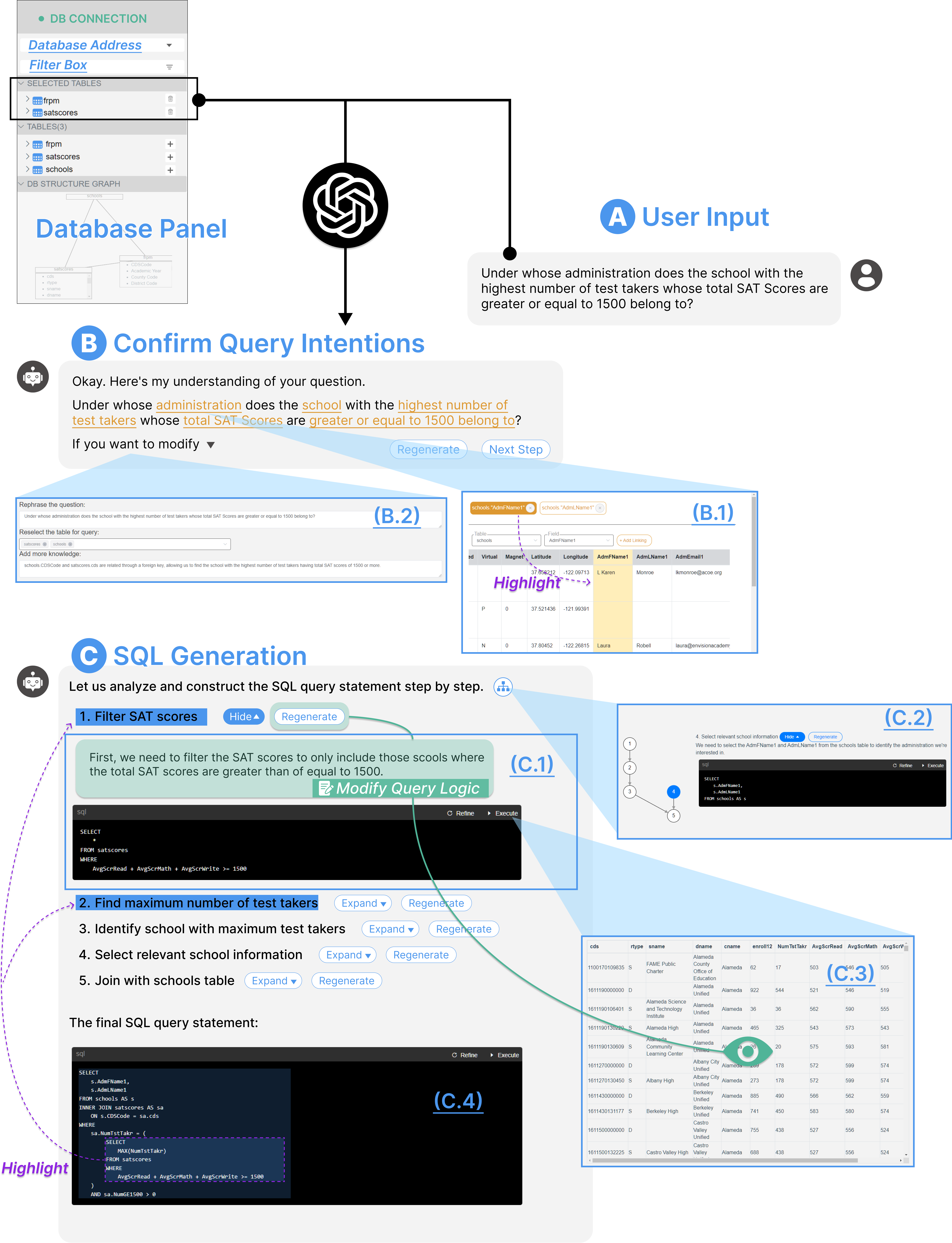}
    \caption{The user interface of \textit{QueryGenie}.}
    \label{fig:CUI}
\end{figure*}

 In the system, the database panel allows users to quickly connect to local databases and intuitively explore their structure. As shown in \autoref{fig:CUI}, users can import a database by selecting its path, and the system will automatically display all tables and fields with keyword-based filtering support. The panel also integrates a UML diagram that visualizes the relationships between tables, making it easier to understand the database schema. In addition, users can select specific tables of interest, and the system will extract their schema information for subsequent interactions with the LLM. This targeted approach prevents irrelevant information from interfering with the LLM's reasoning, improving overall efficiency—especially when working with large databases.

\subsection{Enhanced Conversational Interface}
\par The system leverages an enhanced conversational interface to support transparent and controllable collaboration between user and LLM for data querying. When a user submits an initial query (\autoref{fig:CUI} A), the system provides both the query input and the schema of the selected tables to the LLM, prompting it to establish necessary mappings between natural language entities and database fields. These mappings are then visualized for the user, with matched entities highlighted using yellow underlines (\autoref{fig:CUI} B). Users can click on these highlighted entities to view the corresponding field names and sample data. If any mappings are incorrect, users can either modify the associated fields (\autoref{fig:CUI} B.1) or adjust the query input (\autoref{fig:CUI} B.2) to trigger a new round of schema linking.

\par Once the user confirms that the LLM has accurately understood their intent, the system guides the LLM to generate the query logic through a CoT approach. As shown in \autoref{fig:CUI} (C), the original query is decomposed into five sub-queries, which are displayed in sequence. Users can click the \raisebox{-0.5ex}{\includegraphics[height=2.5ex]{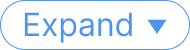}} (Expand) button to view detailed explanations and corresponding sub-SQL statements for each step (\autoref{fig:CUI} C.1). Notably, due to dependencies between reasoning steps—e.g., in \autoref{fig:CUI} (C), Step 2 builds upon the result of Step 1—the query forms a hierarchical tree structure rather than a simple linear sequence. To reflect this structure, we introduce a tree-based visualization (\autoref{fig:CUI} C.2) that illustrates the hierarchical relationships and dependency paths among sub-queries. Users can toggle between the textual and tree views via the \raisebox{-0.8ex}{\includegraphics[height=3ex]{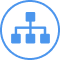}} (Tree) button, enabling a more intuitive understanding of how each step contributes to the overall query logic.

\par Since the system is connected to a database, all SQL statements generated in the conversational interface can be executed directly to support real-time validation. As shown in \autoref{fig:CUI} (C.3), users can click the \raisebox{-0.5ex}{\includegraphics[height=2.5ex]{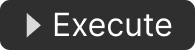}} (Execute) button to view the results of any SQL query. Two common interaction scenarios are supported: (1) If the SQL has syntax errors, users can click the \raisebox{-0.5ex}{\includegraphics[height=2.5ex]{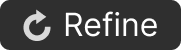}} (Refine) button to prompt the LLM to self-correct and return a revised statement; (2) If the SQL is executable but the results do not align with the intended logic, users can adjust the logic of the relevant step and click the \raisebox{-0.5ex}{\includegraphics[height=2.5ex]{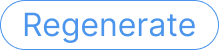}} (Regenerate) button, prompting the LLM to generate an updated SQL based on the modified logic. Given the dependency across sub-queries, editing one step may affect subsequent ones. Although this ``cascade effect'' is logically consistent, users typically do not expect earlier steps to be affected. Therefore, when a node $l_i$ in the logic chain ${l_1 \rightarrow l_2 \rightarrow \cdots \rightarrow l_n}$ is modified, the system preserves all prior nodes ${l_1 \rightarrow \cdots \rightarrow l_{i-1}}$, while regenerating only the subsequent ones ${l_{i+1} \rightarrow \cdots \rightarrow l_n}$.

\par At the bottom of the interface (\autoref{fig:CUI} C.4), the system displays the final SQL statement. To make its structure more understandable, a color-coded visualization is applied: darker colors indicate deeper levels of nesting, and each colored block corresponds to a specific reasoning step. Users can hover over these blocks to view the associated logic, supporting better comprehension and traceability of the final SQL construction process.

\section{Preliminary Evaluation}
\par We conducted a within-subjects user study to preliminarily evaluate the effectiveness of \textit{QueryGenie}. A total of $12$ participants (\textbf{P1--12}, $9$ male, $3$ female; age: $22.19 \pm 3.32$) were asked to complete $6$ data querying tasks using both \textit{QueryGenie} and \textit{Vanna}~\cite{Vanna}, respectively. All tasks were randomly sampled from the BIRD benchmark~\cite{li2024can}. To ensure fairness, both systems used the same underlying LLM, \textit{GPT-4o}. Additionally, the order in which participants used the two systems was counterbalanced to minimize learning effects. During the process, we recorded the completion time and accuracy for each task. After finishing all tasks, participants were invited to take part in a semi-structured interview to provide additional feedback.

\par \textbf{Task Performance.} Participants using \textit{QueryGenie} achieved an overall accuracy of $90.3\%$ ($65/72$), with an average completion time of $203$ seconds per task. In contrast, \textit{Vanna} yielded an accuracy of $66.7\%$ ($48/72$) and an average task time of $130$ seconds. These results suggest that while \textit{QueryGenie} requires more time to complete each task, it enables users to produce significantly more accurate queries. We observed that participants often spent more time using \textit{QueryGenie} to review and understand intermediate reasoning steps generated by the system. Although this process increased task duration, most participants viewed it as a worthwhile trade-off—sacrificing a small amount of time for improved reliability and greater transparency in the querying process.

\par \textbf{Usability Feedback.} In post-study interviews, participants consistently expressed a preference for \textit{QueryGenie}. Many noted that the system ``\textit{integrated multiple useful features}'' ($7/12$), ``\textit{enhanced their understanding of how the LLM generates data queries}'' ($7/12$), and ``\textit{made them feel more confident when completing tasks}'' ($8/12$). The majority of participants ($11/12$) found \textit{QueryGenie} easy to learn and smooth to use. As \textbf{P3} remarked, ``\textit{The system [QueryGenie] introduces some additional features, but I can clearly understand their purpose—they're designed to better support task completion.}'' During the study, we also observed that all participants were able to quickly adapt to the system and became proficient after completing just one or two tasks.

\section{Conclusion}
\par In this study, we proposed an innovative interaction framework and demo system designed to enhance human–LLM collaboration in data querying tasks. Built upon traditional conversational LLMs, our approach integrated visualizations and real-time interactive feedback, enabling users to better understand the model's reasoning process and intervene at critical points. This enhances both the accuracy and transparency of query generation.

\bibliographystyle{ACM-Reference-Format}
\bibliography{sample-base}

\end{document}